\documentclass[prl,twocolumn,amsfonts,showpacs]{revtex4}
\usepackage{graphics,bm,natbib}
\begin{document}
\widetext
\title{Orbital Polarization in Itinerant Magnets}
\author{I. V. Solovyev}
\email[Electronic address: ]{solovyev.igor@nims.go.jp}
\affiliation{Computational Materials Science Center, National Institute for Materials Science,
1-2-1 Sengen, Tsukuba, Ibaraki 305-0047, Japan
}
\date{\today}

\widetext
\begin{abstract}
We propose a parameter-free scheme of calculation of the orbital polarization (OP)
in metals, which starts with the strong-coupling limit for the screened Coulomb
interactions in the random-phase approximation (RPA).
For itinerant magnets,
RPA can be further improved
by restoring
the spin polarization of
the local-spin-density approximation (LSDA)
through the local-field corrections.
The OP is then computed in the static GW approach,
which systematically improves the
orbital magnetization and the magnetic anisotropy energies
in transition-metal and actinide
compounds.
\end{abstract}



\maketitle

  An electron in solid can carry spin ($M_S$) and orbital ($M_L$) magnetic moment.
For weakly correlated systems,
the problem of spin magnetism alone can be formulated in the fully itinerant fashion,
meaning that
the effect of other electrons onto a given one can be described by
an exchange-correlation field (or spin polarization).
The field is typically evaluated in the model of homogeneous electron
gas, in the basis of plane waves, which is a limiting case
of the extended Bloch waves. This constitutes the ground
of the Kohn-Sham (KS) formalism
within LSDA~\cite{KohnSham}, which works exceptionally well
for the magnetic spin properties of many transition-metal and actinide compounds.
They form an extended group of what is
currently called the ``itinerant electron magnets''.

  The orbital magnetism is an \textit{atomic phenomenon}. In the majority of
cases, it is driven by the spin-orbit interaction (SOI), being proportional
to the gradient of the one-electron potential, ${\bf \nabla} \hat{V}$,
which is large only in a small core region close to the atomic nucleus.
Furthermore, the angular momentum
operator, $\hat{L}^z$, does not commute with $\hat{V}$.
Generally, $\hat{L}^z$ is not
an observable quantity, except the same core region, where
$\hat{V}$ is spherically symmetrical. Therefore, it is more natural to
formulate the problem in the basis of Wannier orbitals
$\{ \phi_\alpha \}$ ($\alpha$ being a joint spin-orbital index), localized around
each atomic sites~\cite{condmat05}. Then, the orbital moment
$M_L$$=$${\rm Tr}_{LS} \{ \hat{L}^z \hat{n} \}$ is specified by
the local density matrix
$\hat{n}$$=$$\| n_{\alpha \beta} \|$
(${\rm Tr}_{LS}$ being the trace over spin and orbital variables),
where $n_{\alpha \beta}$$=$$\sum_i n_i d_{\alpha i} d^\dagger_{\beta i}$,
$d_{\alpha i}$$=$$\langle \phi_\alpha | \psi_i \rangle$ is the projection
of KS eigenstate $\psi_i$ onto $\phi_\alpha$,
$n_i$ is the KS occupation number corresponding to the eigenvalue $\varepsilon_i$, and
the joint index $i$ stands for the spin, band, as well as the
position of the ${\bf k}$-point in the first Brillouin zone (BZ).

  In an analogy with the spin polarization
for itinerant magnets, one can think
of an OP: an exchange-correlation field in
KS equations, which couples with $M_L$.
Despite a genuine interest to the problem
and wide perspectives of their potential applications,
the theories of OP in metals
are still in a developing ``semi-empirical'' stage, as they
largely depend on the input parameters, which are typically
chosen to fit the experimental data.
Although majority of researches agree that OP is controlled by intra-atomic
interactions, which are strongly screened in metals,
the details of this screening as well as the form of the OP itself
remains to be a largely unresolved and disputed
problem~\cite{Brooks,Severin,Trygg,PRL98,Yang,CoPt.MAE}.

  Therefore, there are two important questions, which we would like to
address in this work. (i) How the bare on-site interaction
$u_{\alpha \beta \gamma \delta}$$=$$\langle \phi_\alpha \phi_\gamma |
1/r_{12} | \phi_\beta \phi_\delta \rangle$
between $d$- or $f$-electrons
is screened in metals? What is the main mechanism of this screening?
(ii)
Is there any simple and reliable way to evaluate this screening in
\textit{ab~initio} calculations of OP?

  In the atomic limit, the full matrix $\hat{u}$$=$$\| u_{\alpha \beta \gamma \delta} |$
is controlled by a small number of Slater integrals $\{ F^k \}$.
Then, there is an old empirical rule~\cite{empirical}, which states
that in metals, the screening
affects mainly
$F^0$, which contribute to the Coulomb matrix elements
$u_{\alpha \alpha \gamma \gamma}$.
Other Slater integrals, which control the exchange and nonsphericity
of Coulomb interactions
do not change so much.

  First, we argue that the same type of screening can be naturally
obtained
in RPA,
in the fully deterministic fashion. The screened interaction in RPA~\cite{GW,comment.1},
\begin{equation}
\hat{U}(\omega) = \left[1 - \hat{u} \hat{P}(\omega) \right]^{-1}\hat{u},
\label{eqn:URPA}
\end{equation}
depends on
the polarization
$\hat{P}$$=$$\| P_{\alpha \beta \gamma \delta} \|$,
which is treated
in the approximation of noninteracting KS quasiparticles:
\begin{equation}
P^{\rm RPA}_{\alpha \beta \gamma \delta}(\omega) = \sum_{ij}
\frac{(n_i-n_j)
d^\dagger_{\alpha j} d_{\beta i}
d^\dagger_{\gamma i} d_{\delta j}}
{\omega - \varepsilon_j + \varepsilon_i + i\delta (n_i-n_j)}.
\label{eqn:PolarizationPT}
\end{equation}

  The $\omega$-dependence of $\hat{P}$
contributes mainly to the redistribution of the spectral density, whereas
the $\omega$-integrated ground-state properties are controlled by $\hat{U} \equiv \hat{U}(0)$.
Therefore, we consider only the static limit, in which
RPA describes the screening of
$\hat{u}$ caused by the relaxation of $\{ \psi_i \}$
upon removal or addition of an electron
in terms of the perturbation-theory expansion~\cite{PRB05}.

  The simplest \textit{toy model, which illustrates the physics,}
may consist of two spin-polarized bands, formed by
$yz$ (1) and $zx$ (2) orbitals. The model is compatible with the orbital magnetization in the
$\langle 001 \rangle$ direction.
Adopting the following
order of orbitals (within one spin channel):
$\alpha \beta$ ($\gamma \delta$)$=$ $11$, $22$, $12$, and $21$, it is easy to show
that
\begin{equation}
\hat{u} = \left(
\begin{array}{cccc}
 u    & u'   &   0  &  0   \\
 u'   & u    &   0  &  0   \\
 0    & 0    &   0  &  j   \\
 0    & 0    &   j  &  0   \\
\end{array}
\right),
\label{eqn:bareutoymodel}
\end{equation}
where
$u$$=$$F^0$$+$$\frac{4}{49}F^2$$+$$\frac{36}{441}F^4$,
$j$$=$$\frac{3}{49}F^2$$+$$\frac{20}{441}F^4$, and
$u'$$=$$u$$-$$2j$.
Due to the orthogonality of the $yz$ and $zx$ orbitals,
the Coulomb ($\alpha \beta$$=$ $11$, $22$)
and exchange ($\alpha \beta$$=$ $12$, $21$) matrix elements are fully decoupled
from each other.
In order to illustrate the main idea of RPA-screening,
$\hat{P}$ can be taken in the form
$P_{\alpha \beta \gamma \delta}$$=$$P\delta_{\alpha \delta} \delta_{\beta \gamma}$~\cite{comment.4},
which yields:
$U$$=$$[u$$-$$(u^2$$-$$u'^2)P]/[(1$$-$$uP)^2$$-$$(u'P)^2]$,
$U'$$=$$u'/[(1$$-$$uP)^2$$-$$(u'P)^2]$, and
$J$$=$$j/[1$$-$$jP]$.
There is certain hierarchy of bare interactions, and for
many metals
the screening of $u$ and $j$ falls in the strong- and weak-coupling
regime, respectively, so
that $u|P|$$\gg$$1$ while $j|P|$$\ll$$1$~\cite{comment.2}. This yields:
$U$$\simeq$$-$$1/(2P)$$+$$2J$, $U'$$\simeq$$-$$1/(2P)$, and $J$$\simeq$$j$.
Thus, this is the inverse polarization, which plays a role of effective Coulomb interaction
in metals~\cite{PRB05}.
$U$ is strongly screened and \textit{does not depend on the value of bare interaction}.
On the other hand, $J$ is insensitive to the screening.
The multiplier $1/2$ in the expressions for $U$ and $U'$ stands for the orbital
degeneracy. The result can be easily generalized for an
arbitrary number of orbitals $M$
($M$$=$ $5$ and $7$ for $d$- and $f$-electrons, respectively),
which yields $U'$$\simeq$$-$$1/(MP)$~\cite{condmat05}.
In this case, in order to justify the strong-coupling regime, it is sufficient
to have a milder condition,
$uM|P|$$\gg$$1$, which
naturally explains the empirical rule~\cite{empirical}.

\begin{figure}[h!]
\centering \noindent
\resizebox{8.5cm}{!}{\includegraphics{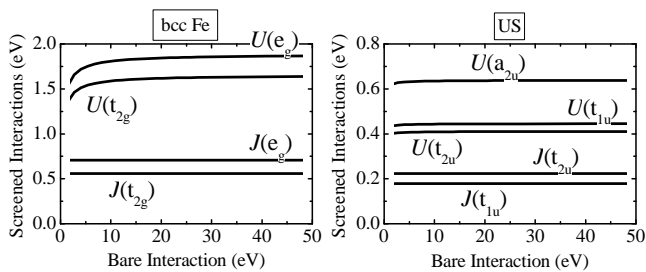}}
\caption{\label{fig.Uversusu}
Effective Coulomb and exchange interactions in RPA
versus bare Slater integral $F^0$ for $3d$-states in bcc Fe
and $5f$-states in uranium sulfide. The symbols denote the
matrix elements corresponding to different representations of the
point group $O_h$. The calculations have been performed in the ferromagnetic
state without spin-orbit coupling.}
\end{figure}

  All these trends are clearly seen in the first-principles calculations
for realistic materials shown in Fig.~\ref{fig.Uversusu}~\cite{comment.3},
where all Slater integrals except $F^0$ were
calculated inside atomic spheres, and $F^0$ was treated as a parameter.
When $F^0$ increases,
the effective interactions quickly reach the asymptotic limit
$F^0$$\rightarrow$$\infty$,
where
$\hat{U}$
is fully determined by details of the electronic structure,
through the polarization $\hat{P}$,
and \textit{do not depend on $F^0$}.
This removes
the main ambiguity with the choice of interaction parameters for metallic compounds.
Since $\hat{P}$ depends on the local
environment in solid, the screened interactions can be different
for different types of Wannier orbitals (e.g., $e_g$ and $t_{2g}$ for $d$-electrons in the
cubic environment).

  Thus, OP in the itinerant magnets
can be naturally evaluated in the
framework of an universal parameter-free scheme based on the
strong coupling limit
for the matrix of effective Coulomb interactions $\hat{U}$.
The
self-energy, incorporating the
effects of OP, can be calculated within static approximation in the GW method~\cite{GW}:
\begin{equation}
\Sigma_{\alpha \beta} =
- \sum_{\gamma \delta} U_{\alpha \delta \gamma \beta} n_{\gamma \delta}.
\label{eqn:SE}
\end{equation}
The proper correction to the KS Hamiltonian in LSDA
is controlled by $\Delta \hat{n}$$=$$\hat{n}$$-$$\frac{1}{2M}\sum_{r=0}^3
{\rm Tr}_{LS}\{ \hat{\sigma}_r \hat{n} \} \hat{\sigma}_r$.
It is obtained after
subtracting the charge ($r$$=$$0$)
and spin ($r$$=$ $1$, $2$, and $3$) density elements of $\hat{n}$, which are already
taken into account in LSDA
($\hat{\sigma}_0$ being the unity matrix, and
$\hat{\sigma}_1$, $\hat{\sigma}_2$, and $\hat{\sigma}_3$ being the Pauli matrices
of the dimension $2M$).
Therefore, in the actual calculations we uses the change of the
self-energy $\Delta \hat{\Sigma}$, which was obtained
after replacing $\hat{n}$ by $\Delta \hat{n}$ in Eq.~(\ref{eqn:SE}).
The problem was solved self-consistently with respect to $\Delta \hat{n}$.

  The validity of the strong-coupling approach is well justified.
So, the effective Coulomb interaction between
$t_{2g}$ electrons in bcc Fe can be estimated
in RPA as 1.50, 1.47, and 1.37 eV for $F^0$$=$ $\infty$, $21$ eV
(the bare Slater integral inside atomic sphere), and $4.5$ eV (the value
obtained in the constraint-LSDA, which includes the screening by the
$sp$-electrons~\cite{PRB05}), respectively.
Thus, even if one takes the lowest estimate $F^0$$=$$4.5$ eV,
the additional approximation $F^0$$\rightarrow$$\infty$ within RPA
would \textit{overestimate} $U$
by less than 10\%.
For $f$-electrons,
this error is even smaller
due to the higher orbital degeneracy.

  However, this is not the main source of the error.
A more fundamental problem is related with the RPA itself,
which typically underestimates the
spin polarization
$\Delta^{\rm RPA}$$=$${\rm Tr}_{LS} \{\hat{\Sigma} \hat{\sigma}_3 \}$,
meaning that
even for the upper limit in RPA, corresponding to $F^0$$\rightarrow$$\infty$,
the effective Coulomb interaction
is \textit{overscreened} and \textit{underestmated}.
For example, had we replaced the spin part of LSDA by the one of RPA,
the spin moment would be underestimated. Obviously, this would destroy
the most attractive point of LSDA for itinerant electron magnets.
Therefore,
there is certain inconsistency in the RPA approach.

  RPA can be improved by introducing the
local-field factor $g$, which incorporates the effects of
exchange-correlation hole for the polarization matrix:
$(\hat{P})^{-1}$$=$$(\hat{P}^{\rm RPA})^{-1}$$-$$\hat{g}$.
Other corrections can
be formally reduced to $\hat{g}$ \cite{Mahan}.
Our goal is to find such a correction to the matrix of effective Coulomb interactions,
which
after substitution in Eq.~(\ref{eqn:SE})
would yield the same
spin polarization as LSDA ($\Delta^{\rm LSDA}$).
In order to do so,
we search $\hat{g}$
in the form of local diagonal matrix:
$g_{\alpha \beta \gamma \delta}$$=$$g \delta_{\alpha \beta} \delta_{\gamma \delta}$.
Then, the asymptotic part of the effective Coulomb interaction
and the self-energy can be easily
recalculated using Eqs.~(\ref{eqn:URPA}) and (\ref{eqn:SE}), respectively, and
the unknown parameter $g$ is obtained
from the condition
${\rm Tr}_{LS} \{\hat{\Sigma} \hat{\sigma}_3 \}$$=$$\Delta^{\rm LSDA}$,
which
is solved self-consistently together with the KS equations.
In the following, this method will be referred to as corrected-RPA (c-RPA).

\begin{figure}[h!]
\centering \noindent
\resizebox{8.5cm}{!}{\includegraphics{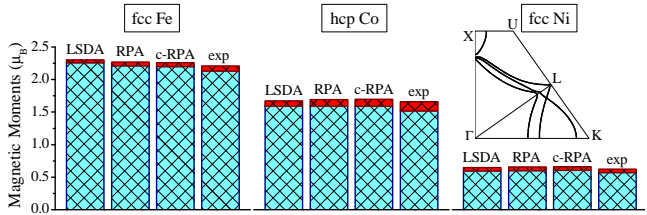}}
\caption{\label{fig.TM}
(Color online) Spin (light blue area), orbital (dark red area), and
total (full hatched area) magnetic
moments in ferromagnetic transition metals.
The experimental data are taken from Ref.~\protect\cite{Trygg}.
The inset shows the Fermi surface
of fcc Ni in the c-RPA approach.}
\end{figure}

  Let us consider first the canonical example of
ferromagnetic transition metals (Fig.~\ref{fig.TM}), where $M_L$ is small
and typically regarded as a small perturbation to the spin-dependent properties.
$M_S$ and $M_L$ can be measured separately using
the x-ray magnetic circular dichroism combined with the spin and orbital sum rules~\cite{XMCD}.
Despite an apparent simplicity,
LSDA encounters a wide spectrum of problems for bcc Fe, hcp Co, and fcc Ni.
We will argue that many of them can be
systematically corrected by applying consequently RPA and c-RPA techniques.
LSDA has certain tendency to overestimate $M_S$ in bcc Fe
and underestimate $M_L$, while RPA and especially c-RPA
substantially improve the LSDA description and yield a good agreement with
the experimental data. The values of $M_S$ ($M_L$)
obtained in LSDA,
RPA, and c-RPA
are $2.26$ ($0.04$), $2.21$ ($0.05$), and $2.20$ ($0.06$) $\mu_B$, respectively,
to be compared with the experimental moments of $2.13$ ($0.08$) $\mu_B$~\cite{Trygg}.
Hcp Co has the largest orbital moment among pure
transition metals ($M_L$$=$$0.14$$\mu_B$),
which is strongly underestimated in LSDA ($M_L$$=$$0.08$$\mu_B$).
The situations is substantially improved in RPA ($M_L$$=$$0.10$$\mu_B$)
and c-RPA ($M_L$$=$$0.11$$\mu_B$).
Fcc Ni is a rare example of ferromagnetic systems
for which $M_L$$=$$0.05$$\mu_B$
is well reproduced already in LSDA.
Both RPA and c-RPA
preserve this good feature of LSDA and do not substantially change $M_L$.
However they do change the electronic structure of fcc Ni. Namely, the form of Fermi surface (FS)
of fcc Ni has been intensively discussed in the context of the
magnetocrystalline anisotropy energy (MAE).
It was argued that the reason why LSDA
fails to reproduce the correct $\langle 111 \rangle$ direction of the magnetization
is related with the second pocket of the FS around the ${\rm X}$-point
of BZ, which is not seen in the experiment~\cite{Yang}.
The experimental FS can be reproduced in the LSDA$+$$U$ approach,
by treating $U$ as an adjustable parameter~\cite{Yang}.
Therefore, it is important that
the same problem can be successfully resolved both in RPA and c-RPA,
\textit{without any adjustable parameters}.
The calculated FS, which reveals only
one pocket around the ${\rm X}$-point, is shown in the inset of
Fig.~\ref{fig.TM}.

\begin{figure}[h!]
\centering \noindent
\resizebox{8.5cm}{!}{\includegraphics{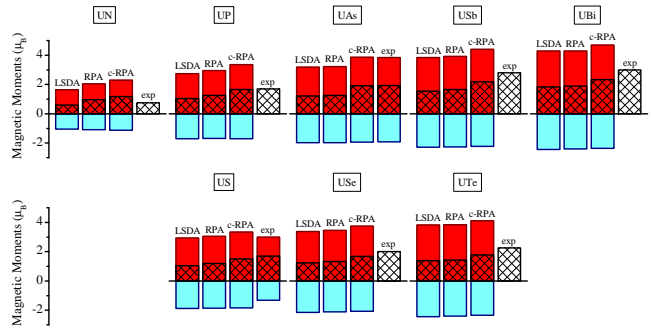}}
\caption{\label{fig.Actinides}
(Color online) Magnetic moments in uranium pnictides (top)
and chalcogenides (bottom). The pnictides (chalcogenides)
have been computed in the type-I antiferromagnetic (ferromagnetic)
structure with $\langle 001 \rangle$ ($\langle 111 \rangle$) direction of the magnetization.
The symbol `exp' shows the results of neutron diffraction, which
were separated into spin and orbital contributions for
US (Ref.~\protect\cite{Severin}) and UAs (Ref.~\protect\cite{Langridge}).
Other notations are the same as in Fig.~\protect\ref{fig.TM}.}
\end{figure}

  The uranium pnictides
(U$X$, where $X$$=$ N, P, As, Sb, and Bi)
and chalcogenides ($X$$=$ S, Se, and Te)
are ones of the most studied actinide compounds. They crystallize in the rock-salt structure.
All chalcogenides are ferromagnets, whereas the pnictides have type-I antiferromagnetic
structure, which may also transform into a multi-${\bf k}$ structure.
The basic difference from the transition metals is that $M_L$ in actinides,
which can be extracted from the analysis of magnetic form factors~\cite{Severin,Langridge},
is very large and typically dominates
over $M_S$. According to the third Hund rule,
$M_S$ and $M_L$ in U$X$ are coupled
antiferromagnetically. As the U-U distance increases,
the U($5f$) states become more localized, and
all magnetic moments
increase monotonously from UN to UBi and from US to UTe (Fig.~\ref{fig.Actinides}).
UN and US are usually classified as itinerant magnets.
However, the role of intra-atomic correlations is expected to increase
for the end-series compounds.
Obviously, the real \textit{ab~initio} scheme does not know whether the
system is itinerant or not. Therefore, it is important to test both RPA and c-RPA
methods for all considered compounds in order to see how they will work for the
materials with the different character of the $5f$-electrons.
The orbital moments are systematically underestimated in LSDA.
The error
is really large so that the total magnetic moments are typically off
the experimental values by
20-50\%.
RPA systematically improves the LSDA description.
However, it is not enough, and for many uranium compounds
it is essential to go beyond RPA.
For these purposes, c-RPA
works exceptionally well and further improves the RPA description.
Particularly, we note an excellent agreement with the experimental data
for $X$$=$ S, P, and As. For the end-series compounds
($X$$=$ Te, Sb, and Bi) the agreement is not so good,
signalling at the necessity of more radical improvements, involving
both orbital \textit{and~spin} polarization of LSDA.
However, even for these complicated systems, c-RPA is a big
step forward over conventional LSDA.

\begin{figure}[h!]
\centering \noindent
\resizebox{8.5cm}{!}{\includegraphics{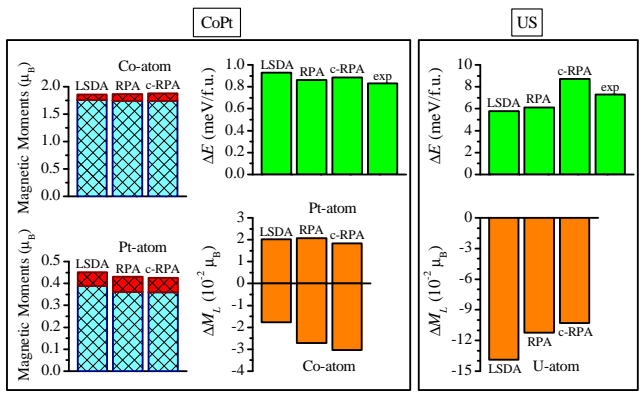}}
\caption{\label{fig.anisotropy}
(Color online) Magnetocrystalline anisotropy energy ($\Delta E$) and
the anisotropy of orbital magnetization ($\Delta M_L$).
For each quantity, the anisotropy is defined as the difference
between values corresponding to the $\langle 100 \rangle$ and $\langle 001 \rangle$ (CoPt),
and $\langle 100 \rangle$ and $\langle 111 \rangle$ (US)
directions of the magnetization. The experimental values are taken from Ref.~\cite{CoPt.MAE}
(CoPt, at 293 K) and Ref.~\cite{US.MAE} (US). For CoPt, the values
of $M_S$ and $M_L$ in the
$\langle 001 \rangle$ direction are shown in the left part of the
figure. Other notations are the same as in Fig.~\protect\ref{fig.TM}.}
\end{figure}

  Finally, let us discuss applications for the MAE. We consider two characteristic
examples: CoPt and US. The ordered tetragonal CoPt alloys is a promising candidate
for magnetic recording applications. An intriguing point is that although LSDA
underestimates $M_L$, MAE is reproduced
surprisingly well (Fig.~\ref{fig.anisotropy})~\cite{CoPt.MAE}.
Therefore, the ``correct'' OP in CoPt should affect only $M_L$.
This requirement is well satisfied both for RPA and c-RPA.
The orbital moments systematically increase in the direction
LSDA$\rightarrow$RPA$\rightarrow$c-RPA to reach
$M_L^{\rm Co}$$=$$0.14\mu_B$ and $M_L^{\rm Pt}$$=$$0.07\mu_B$.
The anisotropy of $M_L$ also
increases (mainly at Co-sites). However, the MAE does not change so much
because of large cancellation of on-site interaction energies associated with
Co and Pt sites~\cite{CoPt.MAE}.

  US has the largest MAE among cubic compounds~\cite{US.MAE}, which is
underestimated in LSDA. The situation is corrected in c-RPA, at least
qualitatively. It is curious that MAE ``anticorrelate'' with the
anisotropy of orbital magnetization, which \textit{decreases} in the
direction LSDA$\rightarrow$RPA$\rightarrow$c-RPA.
However, this is not surprising, because in cubic compounds, MAE is the
forth order effect with respect to SOI. Therefore,
there is no
simple relation between $\Delta E$ and $\Delta M_L$ and the main correction to MAE
in c-RPA comes from the change of the on-site interaction energy.

  In summary, we have argued that
the problem of OP in metals can be naturally formulated
``from the first principles'',
by considering
the strong-coupling limit for the screened Coulomb interactions.
In the present work, the screened $\hat{U}$ was computed only once: in LSDA and
without SOI.
An important extension would be a self-consistent determination of $\hat{U}$,
which would incorporate the effects of OP into the screening.
(i) It could improve
the description of some itinerant actinide compounds
(e.g., UN)
for which the spin polarization is small, and the screening is strongly
influenced by SOI. (ii) Since the OP affects
the KS eigenvalues
$\{ \varepsilon_i \}$, which stand
in the denominator
of the polarization matrix (\ref{eqn:PolarizationPT}), the screening
is expected to decrease.
This could extend the applicability of the proposed
method for materials
with more localized $5f$- and $4f$-electrons.


\begin{thebibliography}{10}
\bibitem{comment.4}
In this approximation,
each band is formed by either $yz$ or $zx$ orbitals, which do not mix
in KS eigenfunctions.
\bibitem{comment.2}
Typical values of parameters for transition metals are
$u$$\sim$$u'$$\sim$25-30 eV, $j$$\sim$1 eV, and $|P|$$\sim$0.2-0.3 eV$^{-1}$.
\end{thebibliography}
\end{document}